# Unconscious Gender Bias in Academia: from PhD Students to Professors

K. Poppenhaeger[1, 2, a)]

[1]*Queen's University Belfast, Astrophysics Research Centre, University Road, BT7 1NN Belfast, United Kingdom*
[2]*Harvard-Smithsonian Center for Astrophysics, 60 Garden Street, Cambridge, MA 02138, USA*

[a)]Corresponding author: k.poppenhaeger@qub.ac.uk

**Abstract.** In an academic landscape where female physicists are still strongly underrepresented, underlying causes like unconscious gender bias deserve specific attention. Members of academia are often not aware of their intrinsic, hence unconscious, biases; this can have negative effects on students and staff at all career levels. At the Queen's University Belfast, I have developed and conducted a workshop on unconscious gender bias awareness at the School of Mathematics and Physics. The first installment of the workshop was attended by 63 members of the School, among them 26 academic staff (lecturer level and above). Participants attended an informational talk followed by a discussion session, and then took part in the Harvard Implicit Association Test for association of gender with science. The participants self-reported their results and their previous expectations, followed by a group discussion. Here I present the observed magnitude of unconscious gender bias and summarise the discussion points of the participants. The outcomes that bias can have on the success of physics students as well as the careers of physicists in an academic context will be highlighted. Putting the results into context, I discuss steps forward to make physics a level playing field for all genders.

## INTRODUCTION TO UNCONSCIOUS BIAS

The definition of bias is a positive or negative unconscious belief about a particular category of people. This allows quick, but sometimes inaccurate, processing of information. It often conflicts with consciously held attitudes. Over time, biases can change based on experience and exposure. Some examples are: On average, both men and women underestimate the contributions of women. Similarly, on average both whites and people of colour underestimate the contributions of people of colour. Biases are not the same as discrimination; discrimination can happen if a person actually acts on their biases. However, if someone is aware of their biases, they can consciously decide if they act on them. If biases go unchecked, they can have multiple detrimental effects for groups against which negative biases exist, for example in performance evaluations, hiring, and career progression.

Unconcious gender bias can have a significant effect on how students and their proficiency are evaluated by academic staff. Moss-Racusin et al. (2012) studied if professors in STEM fields rate identical student CVs differently if the CV lists a female or a male first name. On average, the professors would rate the "male student" significantly more positively on the aspects of competency and hireability, would offer to mentor the student more often, and offer them a 10% higher salary on average. The gender of the professor did not influence how they responded on average, showing that unconscious biases about a group can also be held by members of that group.

Unconscious gender bias can also produce significant differences in how male and female academic staff are evaluated by students. MacNell et al. (2015) conducted a study to quantify outcomes of gender bias in student evaluations by taking advantage of online teaching methods in order to have the students be "blind" to the actual gender of their academic teacher. If students thought they were taught by a woman, they gave significantly lower teaching quality ratings than when they thought they were taught by a man.

Similar trends have been found with respect to unconscious racial bias, both with respects to teachers having negative biases against students of colour (McKown and Weinstein, 2002) and students as well as peers against academic teachers of colour (Huston, 2005).

# DATA FROM PHYSICS AND MATHEMATICS AT QUEEN'S UNIVERSITY BELFAST

I conducted a workshop on unconscious bias awareness at the School of Mathematics and Physics of Queen's University Belfast (QUB) in 2016. This workshop consisted of a talk I gave to the group to introduce them to the concept of unconscious bias, a short discussion session after the talk, a self-assessment online test about unconscious gender bias, and answering an anonymous feedback questionnaire. For the self-assessment test, the Harvard Implicit Association test was used (Greenwald et al. 2003). This is a test where one needs to quickly sort male and female words as well as words that correspond to "Natural Sciences" or "Arts and Humanities"; how accurately one performs under time pressure is used as a measure for how natural the tested gender-science associations feel to the person taking the test. Non-binary genders and their perceived association with science are not evaluated in this particular test. This test works well to determine the biases of groups of people, but is not suitable as a precise test for the bias of an individual person, since the scatter on individual repeatability is rather large.

## Results from the Workshop

My workshop at QUB was attended by 63 people from the School of Mathematics and Physics, out of which 54 completed the full workshop. The vast majority of participants were from physics institutes (93%). Out of all participants, 17 were PhD students, and 28 were academic staff members (this UK term corresponds to international career levels of Assistant Professor up to Full Professor). The remainder were postdoctoral research assistants (7) or support staff and other categories (2). 38 attendees were male, 16 were female, and none of the attendees self-reported a non-binary gender.

The test evaluated how strongly one associates the male or female gender with natural sciences; possible results ranged from a strong male-plus-science association to a strong female-plus-science association. The participants showed an overall tendency to associate the male gender more strongly with science than the female gender (Fig. 1a). The largest group of attendees had a neutral association of gender with natural sciences; however, the second and third largest groups had moderate and weak associations of the male gender with natural sciences, respectively. Much smaller groups had unconscious associations of the female gender with natural sciences. I also asked the participants if they were surprised by the result of the test. Overall, the participants' expectation and the actual result were similar: the largest group of participants reported that their test result matched their expectations, while the groups that reported a more female-skewing or male-skewing test result than expected were of almost similar size (Fig. 1b).

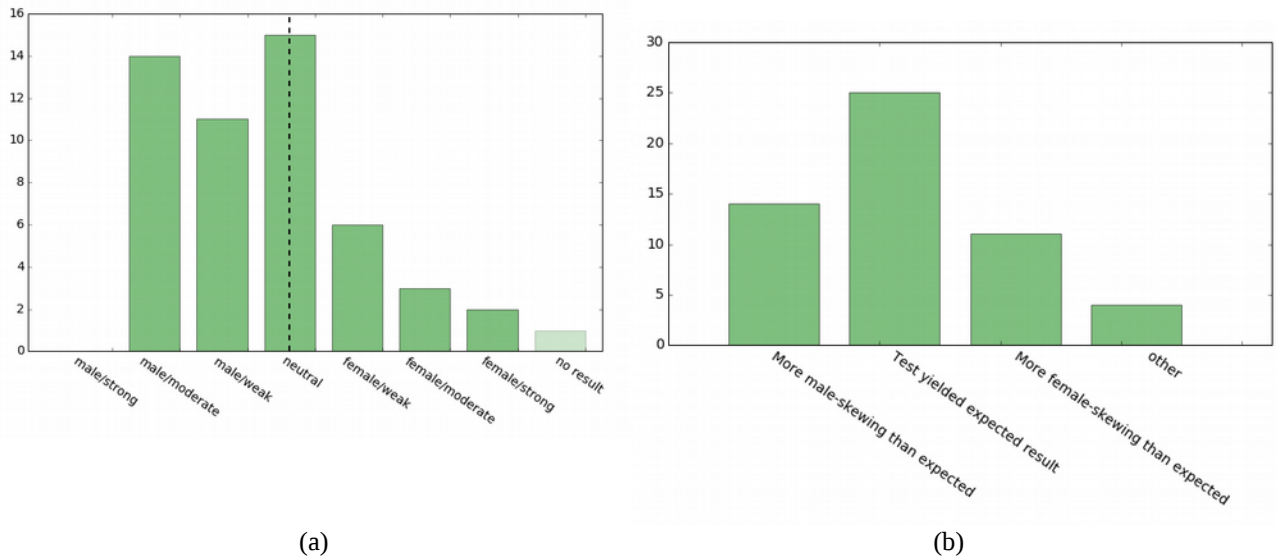

(a)          (b)

**FIGURE 1.** (a) Association of male or female gender with natural sciences in the Harvard Implicit Association Test as reported by the participants of the workshop. (b) Expectations of the attendees about their own gender-science bias.

The results achieved on the test were different for the female and male participant groups; overall, female participants showed less tendency to associate the male gender with science (Fig. 2a). Specifically, the group of female participants displayed a distribution that peaked at a neutral association of gender with natural sciences, and displayed a similar distribution to male-plus-science associations and female-plus-science associations. In contrast, the male attendees displayed a distribution that peaked at a moderate association of the male gender with science, with a decrease in numbers towards neutral and female-plus-science associations. The gender-split in the results here is rather surprising: in larger studies, authors usually find that both female and male participants of the studies display negative biases against women in natural sciences, see for example Moss-Racusin et al (2012). However, in this workshop we specifically only tested for unconscious associations, not how people would act upon their biases if presented with a more realistic situation as in the study by Moss-Racusin and co-authors, which may explain the neutral result displayed by the female participants of this workshop.

I compared if the average age of the participants had an effect on the test result. Using the two largest participant groups, PhD students were used as the on average younger test group, and academic staff as the on average older test group. Both groups show a tendency to associate the male gender with science, academic staff slightly less so than PhD students (Fig. 2b). While one might expect that the younger generation might be less influenced by gender stereotypes and their perceived fit to natural sciences due to growing up in more modern times, no obvious support for such a hypothesis is found in this data set.

## Comments from Participants

Comments were collected from participants at the QUB workshop, as well as from another round of the workshop I gave at the Spring Meeting of the Institute of Physics in Dublin 2017. Several female participants mentioned that they felt validated in their experiences through the data presented in the talk. Other participants, especially male ones, were surprised at the magnitude of the effect observed in student evaluations reported by McNell (2015). One male participant noted that this was an eye-opening experience for him, and he could relate to the impact of unconscious gender bias through his own experience of negative bias against older scientists in hiring processes. In a more formalized manner through a questionnaire, I collected feedback from the participants about how useful they found the workshop. Overall, the participants generally found the workshop useful, with 90% rating the events as useful or very useful and stated that they learned something new during the event (74%).

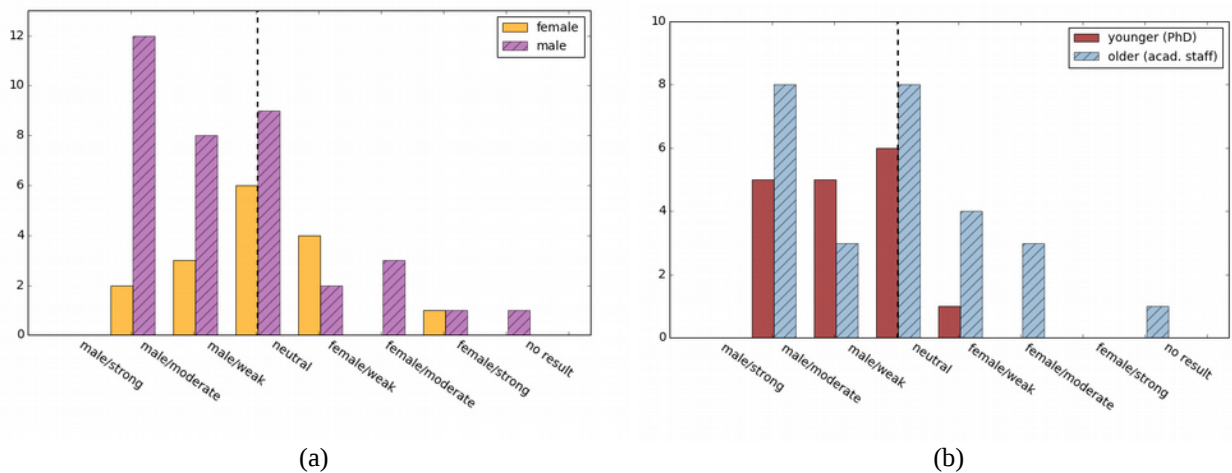

(a)  (b)

**FIGURE 2.** (a) Association of male or female gender with natural sciences, split up by gender of workshop participants. (b) Association of male or female gender with natural sciences, split up by academic status of workshop participants (PhD students versus academic staff).

# CONCLUSIONS

The collected data shows that unconscious gender bias is present in members of the School of Mathematics and Physics at QUB at all academic levels, ranging from PhD students to academic members of staff. Given previous studies, it would actually have been surprising if unconscious gender bias had not been detected. Interestingly, the female participants of the workshop did not display an unconscious bias against women in natural sciences, contrary to peer-reviewed studies finding that both men and women tend to have anti-female biases. If this effect can be confirmed in other academic settings, women might be fairer in evaluating candidates for hiring and academic career progression. In order not to overburden female academics, one should consider giving other workload reductions to female academics if they are asked to serve on such committees more often. It is furthermore important to emphasise that unconscious biases of any kind do not necessarily need to amount to discrimination. Only when unconscious biases are allowed to influence our actions do they actually lead to discrimination. One way to decrease the influence of biases is to have policies in place that govern important steps of education and career progression, for example student admissions, job interviews, and promotions. Another important point is to be aware of one's biases, so that one can be vigilant and reflect on the reasons for the decision one makes. Given that a large majority of participants of the workshop found it useful and felt they have learned something new, bias awareness workshops and discussions may be a suitable tool to help people to decrease the influence of their biases on their actions.

# ACKNOWLEDGMENTS

This contribution is based on work undertaken by Dr Poppenhaeger as part of the Postgraduate Certificate in Higher Education (PGCHET). The relevant modules of the PGCHET were taught at Queen's University Belfast by Dr Joseph Allen in 2016/17.

Dr Poppenhaeger's participation in ICWIP conference was funded by the School of Mathematics and Physics at Queen's University Belfast.

# REFERENCES

1. Moss-Racusin, C.A., Dovidio, J.F., Brescoll, V.L., Graham, M.J. and Handelsman, J. (2012), 'Science faculty's subtle gender biases favor male students', Proceedings of the National Academy of Sciences of the United States of America, 109(41), pp16474-16479.
2. MacNell, L., Driscoll, A. and Hunt, N.H. (2015), 'What's in a Name: Exposing Gender Bias in Student Ratings of Teaching', Innovative Higher Education, 40(4), pp291-303.
3. McKown, C. and Weinstein, R.S. (2002), 'Modeling the Role of Child Ethnicity and Gender in Children's Differential Response to Teacher Expectations', Journal of Applied Social Psychology, 32(1), pp159-184.
4. Huston, T. A. (2005), 'Race and Gender Bias in Higher Education: Could Faculty Course Evaluations Impede Further Progress Toward Parity?', Seattle Journal for Social Justice, 4(2), pp591-611.
5. Greenwald, A. G., Nosek, B. A., and Banaji, M. R. (2003), 'Understanding and using the Implicit Association Test: I. An improved scoring algorithm.' Journal of Personality and Social Psychology, 85(2), 197-216.